\documentclass[12pt,preprint]{aastex}

\newcommand{\der}[2][\;\;]{\ensuremath{ \frac{d{#1}}{d{#2}} }}

\newcommand{\dpar}[2][\;\;]{\ensuremath{ \frac{\partial{#1}}{\partial{#2}} }}

\newcommand{\iint}[1]{\int\limits_{~~#1}\!\!\!\!\int}

\newcommand{\grad}{\bvec{\nabla}}

\newcommand{\e}{{\rm e}}
\newcommand{\D}{\displaystyle}

\newcommand{\bvec}[1]{{\mbox{{\boldmath$#1$}}}} 
\newcommand{\unitv}[1]{\bvec{\hat{#1}}}

\newcommand{\eqnref}[1]{(\ref{#1})}

\begin{document}

\title{Equilibrium models of coronal loops that involve curvature and buoyancy}

\author{Bradley W. Hindman}
\affil{JILA and Department of Astrophysical and Planetary Sciences,
University of Colorado, Boulder, CO~80309-0440, USA}

\author{Rekha Jain}
\affil{School of Mathematics \& Statistics, University of Sheffield, Sheffield S3 7RH, UK}

\email{hindman@solarz.colorado.edu}


\begin{abstract}

We construct magnetostatic models of coronal loops in which the thermodynamics
of the loop is fully consistent with the shape and geometry of the loop. This is
achieved by treating the loop as a thin, compact, magnetic fibril that is a small
departure from a force-free state. The density along the loop is related to the
loop's curvature by requiring that the Lorentz force arising from this deviation
is balanced by buoyancy. This equilibrium, coupled with hydrostatic balance and
the ideal gas law, then connects the temperature of the loop with the curvature
of the loop without resorting to a detailed treatment of heating and cooling.
We present two example solutions: one with a spatially invariant magnetic Bond number
(the dimensionless ratio of buoyancy to Lorentz forces) and the other with a
constant radius of curvature of the loop's axis. We find that the density and
temperature profiles are quite sensitive to curvature variations along the loop,
even for loops with similar aspect ratios.
\end{abstract}

\keywords{MHD --- Sun: Corona --- Sun: magnetic fields}


\section{Introduction}
\label{sec:introduction}
\setcounter{equation}{0}

In EUV images of the solar corona, coronal loops appear as long graceful arcs of
bright plasma that trace magnetic field lines through the atmosphere. These loops
can be preferentially illuminated because localized heating and inefficient cross-field
diffusion lead to hot plasma spreading along individual field lines. Despite the
obvious magnetic nature of these structures, it has proven challenging to measure
through spectroscopic means the magnetic-field strength within the corona. Of course
measuring the magnetic-field strength is equivalent to measuring the energy density,
and is therefore a key constraint in the modeling of energetic and eruptive phenomena
such as flares and CMEs.

Coronal loops are sometimes observed to vacillate back and forth with a regular
frequency. The identification of these oscillations as resonant MHD kink waves,
trapped between the loop's footpoints \citep[e.g., ][]{Aschwanden:1999, Aschwanden:2011-apj,
Nakariakov:1999, Verwichte:2004}, launched the field of coronal loop seismology.
Coronal seismology promises the opportunity to measure the magnetic-field strength along
the loop through inversion of the kink-mode eigenfrequencies \citep[e.g., ][]{Jain:2012}.
However, before such inversions can be performed the ability to construct stable,
curved coronal loop models with realistic density and magnetic profiles is needed.

The state of the art in the modeling of static coronal loops is nonlinear force-free
field (NLFFF) models. Typically such models have used vector magnetograph measurements
in the photosphere as an observational constraint. A substantial weakness to
such an approach is that the force-free assumption is rather inappropriate within the
chromosphere and the low corona \citep{Metcalf:1995}; thus, the region in which the
force-free assumption is valid and the region in which the observational constraint
is applied are disjoint. This leads to a substantial mismatch between the model field
and the actual magnetic field \citep{DeRosa:2009}. More recent work has partially
overcome this difficulty by applying additional constraints higher in the corona.
These constraints have taken the form of EUV images of bright coronal loops from
instruments such as STEREO and the {\sl Atmospheric Imaging Assembly} (AIA). Initially,
multiple such EUV images were used, each taken from a different vantage point either
using simultaneous images from the two STEREO spacecraft or using solar rotation to
view the presumably static magnetic structure from different angles. Solar stereoscopy
was then used to reconstruct the 3-D curve traced by a loop
\citep[see the reviews of][]{ Wiegelmann:2009, Aschwanden:2011-LRSP}. However, recent
attempts to deduce the 3-D shape of a loop from just a single high-resolution EUV image
(such as those from AIA) and a coeval photospheric magnetogram have had intriguing
success \citep{Aschwanden:2013}.

Despite the achievements that the NLFFF models have made in reconstructing the geometry
of the magnetic field, the force-free assumption decouples the thermodynamic variables
from the magnetic field. Certainly, if the dynamic timescales are significantly shorter
than the cooling times, hydrostatic balance along field lines is still valid. However,
deducing the mass density and gas pressure requires either the specification of the
temperature by fiat, or the inclusion of an energy equation that models the heating and
cooling of the loop \citep[e.g., ][]{Aschwanden:2001, Cargill:2004, Winebarger:2004,
Patsourakos:2008, Reale:2002, Reale:2010}. Here we constrain the mass density
within the loop in a different manner. In essence, we find an equilibrium solution by
considering deviations from a force-free state. This deviation only occurs within the
loop and appears as a change in the field strength without a concomitant change in the
field's direction. Such a field strength perturbation produces both magnetic buoyancy
and a small Lorentz force, both of which depend intimately on the shape or geometry of the
loop and which oppose each other in equilibrium. The establishment of this equilibrium 
requires that the mass along the loop redistributes itself with a timescale shorter than
the cooling time. Thus, accounting for the buoyancy of the loop allows one to directly
connect the mass density and other thermodynamic variables
(such as the temperature) to the shape of the loop and the magnetic-field strength within
the loop. We will show that the temperature profile of the loop is not a function that
can be freely specified, but instead has a functional form that is a direct consequence
of the geometry of the loop.

Our goal here is to self-consistently include curvature and buoyancy in the equilibria
of coronal loops and to develop models of the temperature, density, and field strength
with the geometry of the loop as the primary input.
In order to permit analytic solutions we treat coronal loops as slender magnetic
fibrils and adopt the thin-flux-tube approximation when deriving the force balance.
The balance of forces is characterized by a magnetic Bond number which is the
dimensionless ratio of the buoyancy force to the Lorentz force. The shape
and curvature of the loop is succinctly expressed in terms of the magnetic
Bond number, which may be a function of position along the loop. In \S2 we
will derive the equation that describes the balance of forces and from this equation
we identify the magnetic Bond number. Then, assuming that the shape of the loop
is provided by observations, we derive the temperature and mass density profiles
that are consistent with this shape. In \S3 we present a simple equilibrium
solution for an embedded fibril which has a uniform magnetic Bond number.
We discuss the atmospheric properties that are consistent with a constant
magnetic Bond number and derive the resulting coronal magnetic field. In \S4
we demonstrate another simple solution corresponding to a semi-circular fibril
with a uniform radius of curvature. Finally, in \S5 we discuss the implications
of our findings and summarize the results.


\section{Force Balance for a Thin Loop}
\label{sec:Forces}

We will model a coronal loop as a curved, magnetic fibril embedded in a larger
coronal magnetic structure. We will further assume that the fibril is compact and
thin. By {\sl thin} we mean that the radius of the fibril is small compared to any
other relevant length scale. Even though the thin-tube approximation may not apply
to all coronal loops, it is an appropriate approximation for many. Recent
high-spatial-resolution observations of the solar corona in the Fe XIII 19.5 nm
line by the Hi-C instrument have enabled a resolution of about 150 km, sufficient
to resolve the cross-section of most coronal loops. Using these observations,
\cite{Brooks:2013} examined brightness cross-sections for 91 loops in the solar
corona and found a distribution of radii sharply peaked at 270 km. By examining
the pixel-to-pixel brightness fluctuations across loop cross-sections, \cite{Peter:2013}
have argued that coronal loops are unlikely to be structured on a finer unresolved
spatial scale. Since the corona's pressure and density scale heights are generally
a hundred times larger than this spatial scale and the lengths of loops a thousand
times larger, the thin-flux-tube approximation appears to be quite relevant for a
substantial fraction of coronal loops.

We also assume that the external corona is magnetically dominated and its magnetic
field is a force-free field. We adopt the notation that quantities
evaluated in the external corona have a subscript `e', while those within the fibril
lack a subscript. Thus, the external magnetic field is $\bvec{B}_\e$; whereas, the
internal magnetic field of the fibril is just $\bvec{B}$. One way to envision the
fibril is to select a bundle of field lines within the coronal magnetic field and
uniformly increase or decrease the field strength within that bundle by a constant
factor, $B^2 = \left(1+\alpha\right) B_\e^2$. We call the constant $\alpha$ the
field-strength deviation and it can be positive or negative depending on whether the fibril
is strongly or weakly magnetized compared to its surroundings. We consider only loops
that are small deviations from the force-free state. Thus, the
field-strength deviation will be a small quantity, $\left|\alpha\right| <<1$. The
constant $\alpha$ was chosen such that it represents the constant of proportionality
between the exterior magnetic pressure and the magnetic-pressure contrast (the difference
in the magnetic pressure between the inside and outside of the fibril),

\begin{equation} \label{eqn:MatchedFlare}
	\frac{\Delta B^2}{8\pi} \equiv \frac{B^2-B_\e^2}{8\pi} = \alpha \, \frac{B_\e^2}{8\pi} \; .
\end{equation}

\noindent The field-strength deviation $\alpha$ must be constant along the fibril. Otherwise
the fibril and surrounding corona would not have a common flux surface where they
join. Since the internal magnetic field is proportional to the external magnetic
field, the internal field is also force free. However, because of the discontinuity
in the field strength at the edge of the fibril, there exists a current sheath that
surrounds the fibril.

We neglect the spherical geometry of the solar atmosphere and assume that the corona
can be treated as a plane-parallel atmosphere with constant gravity $\bvec{g}$. We
employ a Cartesian coordinate system, with the $x$--$y$ plane corresponding to the
photosphere and the $z$ coordinate increasing upwards (i.e., $\bvec{g} = -g \unitv{z}$).
We restrict our attention to coronal loops that are symmetric about the origin and
that are confined to the $x$--$z$ plane. Such loops lack torsion.

In addition to the Cartesian coordinate system, both within the fibril and within
the external corona we will employ the local Frenet coordinates for a field line
(illustrated in Figure 1). The direction tangent to the magnetic field will be
denoted with the unit vector $\unitv{s}$ (thus $\bvec{B}_\e = B_\e \unitv{s}$),
and the longitudinal coordinate $s$ is the pathlength along a field line measured
from the photosphere ($s=0$ corresponds to the footpoint intersecting the photosphere
in the region $x<0$). The curvature vector for the field line is indicated by
$\bvec{k}$, and points in the direction of the principle normal $\unitv{k}$ with
a modulus equal to the reciprocal of the local radius of curvature $R$ of the field
line. The direction of the unit vector in the binormal direction will be indicated
with $\unitv{q}$ and the torsion of the field line with the variable $\tau$. In
the equilibrium considered here, the loop itself lacks torsion, $\tau=0$, and its
binormal uniformly points in the $y$-direction, $\unitv{q} = \unitv{y}$. However,
the exterior coronal magnetic field may be 3-D (with spatial symmetries near the
loop). Therefore, for completeness and to aid follow-up work we consider the more
general case for the moment and specialize only as necessary. The standard geometrical
relations between these coordinate vectors, i.e., the Frenet-Serret formulae, are
given below for reference, assuming that the position vector of a field line is
given by $\bvec{r}(s,t)$,

\begin{eqnarray}
\label{eqn:s_hat}
	\unitv{s} &\equiv& \dpar[\bvec{r}]{s} \; ,
\\ \nonumber \\
\label{eqn:k_hat}
	\bvec{k} &\equiv& \dpar[\unitv{s}]{s} = \frac{\unitv{k}}{R} \; ,
\\ \nonumber \\
	\unitv{q} &\equiv& \unitv{s} \times \unitv{k} \; ,
\\ \nonumber \\
	\dpar[\unitv{k}]{s} &=& -\frac{\unitv{s}}{R} + \tau \unitv{q} \; ,
\\ \nonumber \\
	\dpar[\unitv{q}]{s} &=& -\tau\unitv{k} \; .
\end{eqnarray}

\noindent Notice the lack of `e' subscripts on the Frenet vectors. For a sufficiently
thin tube the Frenet vectors will be nearly constant across the fibril with the same
value as the surrounding corona. Therefore, we avoid appending subscripted labels
to the Frenet vectors of the external field to simplify notation and to emphasize
that the field lines have the same direction and shape inside and immediately outside
the fibril.


\subsection{Cross-Sectional Averaging of the MHD Momentum Equation}
\label{subsec:CS_Ave}

The forces acting on an isolated, thin, magnetic fibril embedded in a {\sl field-free}
atmosphere have been previously derived by a variety of authors \citep{Spruit:1981,
Choudhuri:1990, Cheng:1992} through averaging of the MHD momentum equation over the
cross-sectional area of the tube. Here we rederive the equilibrium forces in the
presence of a {\sl magnetized} external atmosphere. Following \cite{Spruit:1981},
we begin by averaging the MHD momentum equation over the cross-section of the fibril,
with the goal of deriving an equation which describes the force per unit length along
the fibril. In equilibrium, there will be a balance of these forces, and this balance
will specify the shape of the fibril and its thermodynamic properties.

Let $A(s)$ be a cross-sectional surface of the fibril at the location $s$ that is
everywhere perpendicular to the magnetic field (i.e., perpendicular to $\unitv{s}$)
and let $da$ be a differential area of this surface. We define the cross-sectional
average of a general quantity $f$ in the natural way,

\begin{equation} \label{eqn:AveDef}
	\bar{f}(s) \equiv \frac{1}{A(s)} \int\limits_{\! A(s)}\!\!\int f \, da \; .
\end{equation}

\noindent We write the MHD momentum equation using a formulation of the Lorentz force
that directly acknowledges that the magnetic force is transverse to the field itself,

\begin{equation} \label{eqn:MHDmomentum}
	\rho \frac{D\bvec{v}}{Dt} = -\grad{P} - \grad_\perp \left(\frac{B^2}{8\pi}\right) + \frac{B^2}{4\pi}\bvec{k} + \bvec{g}\rho \; ,
\end{equation}

\noindent where

\begin{equation}
	\grad_\perp \equiv \unitv{k}(\unitv{k} \cdot \grad) + \unitv{q}(\unitv{q} \cdot \grad).
\end{equation}

\noindent In equation~\eqnref{eqn:MHDmomentum}, the gas pressure, magnetic-field strength,
and mass density within the loop are $P$, $B$, and $\rho$, respectively and $D/Dt$ denotes
the Lagrangian time derivative.

We average equation~\eqnref{eqn:MHDmomentum} over the cross-section $A$, and seek a
static solution by setting the acceleration to zero,

\begin{equation}
	-\frac{1}{A} \iint{A} \left[ \grad P + \grad_\perp \left( \frac{B^2}{8\pi} \right) \right] \, da
		+ \frac{ \overline{B^2 \bvec{k}} }{4\pi} + \bvec{g} \bar{\rho}  = 0 \; .
\end{equation}

\noindent The two terms involving gradients can be expressed as contour integrals
around $\partial A$, the boundary of $A$, by using the 2-D form of the divergence theorem
appropriate for integration over an open surface,

\begin{equation} \label{eqn:AveMomentum}
	-\frac{1}{A} \oint\limits_{\partial A} \left(P + \frac{B^2}{8\pi} \right) \unitv{n} \, dl
		- \overline{\dpar[P]{s}\unitv{s}} + \frac{ \overline{B^2 \bvec{k}} }{4\pi} + \bvec{g} \bar{\rho} = 0 \; .
\end{equation}

\noindent The unit vector $\unitv{n}$ is the outward normal to the tube's bounding
surface. The differential $dl$ is the differential pathlength around $\partial A$.
In deriving this equation, we have decomposed the gradient of the gas pressure into
longitudinal and transverse components,

\begin{equation}
	\grad P = \dpar[P]{s} \, \unitv{s} + \grad_\perp P \; .
\end{equation}

The total pressure must be continuous across the loop's bounding surface. Therefore, on
$\partial A$

\begin{equation} \label{eqn:PressureCont}
	P + \frac{B^2}{8\pi} = P_\e + \frac{B_\e^2}{8\pi} \; ,
\end{equation}

\noindent and the pressures that appear within the integrand of the contour integral
in equation~\eqnref{eqn:AveMomentum} may be replaced with those from the external
fluid,

\begin{equation} \label{eqn:ForceBalance}
	-\frac{1}{A} \oint\limits_{\partial A} \left(P_\e + \frac{B_\e^2}{8\pi} \right) \unitv{n} \, dl
		- \overline{\dpar[P]{s}\unitv{s}} + \frac{ \overline{B^2 \bvec{k}} }{4\pi} + \bvec{g} \bar{\rho} = 0 \; .
\end{equation}

Buoyancy is the sum of the gravity acting on a body and the net gas-pressure force acting
on the outer surface of the body. In this case, we also must consider the effects of the
external magnetic pressure and magnetic tension. We can construct an extension of Archimedes'
principle that is relevant for our problem by performing a similar cross-sectional average
of the MHD momentum equation in the exterior fluid. This requires that we assume that the
mathematical form of the external pressures and density can be analytically continued inside
the loop. The transverse component of the resulting equation provides an expression for the
net external pressure force,

\begin{equation} \label{eqn:ExtPressure}
	-\frac{1}{A} \oint\limits_{\partial A} \left(P_\e + \frac{B_\e^2}{8\pi} \right) \unitv{n} \, dl =
		-\frac{ \overline{B_\e^2 \bvec{k}} }{4\pi} - \overline{ \rho_\e \bvec{g}_\perp } \; ,
\end{equation}

\noindent where $\bvec{g}_\perp = \bvec{g} - \left(\bvec{g}\cdot\unitv{s}\right) \unitv{s}$
is the component of $\bvec{g}$ perpendicular to $\unitv{s}$. Since we have assumed that the
tube is thin, to lowest order in the radius of the tube, the tangent vector and curvature
vector are constant across a cross-section. Thus, the previous equation can be rewritten,

\begin{equation} \label{eqn:force_thin}
	-\frac{1}{A} \oint\limits_{\partial A} \left(P_\e + \frac{B_\e^2}{8\pi} \right) \unitv{n} \, dl =
		-\frac{ \overline{B_\e^2} }{4\pi} \, \bvec{k} - \bar{\rho}_\e \bvec{g}_\perp \; .
\end{equation}

This expression can be used to eliminate the external pressure integrals in equation~\eqnref{eqn:ForceBalance}.
If we once again assume that the tube is thin and the Frenet vectors do not vary significantly
over the tube's cross-section, we can replace the cross-sectional averages of the densities and
pressures with their axial values to obtain

\begin{equation} \label{eqn:MeanForces}
	\left[ -\dpar[P]{s} + g_\parallel \rho \right] \unitv{s} + 
		\frac{B^2-B_\e^2}{4\pi} \, \bvec{k} + \left( \rho - \rho_\e\right) \bvec{g}_\perp = 0 \; .
\end{equation}

\noindent where $g_\parallel = \bvec{g} \cdot \unitv{s}$ is the tangential component of
gravity. For simplicity we have labeled the axial values without accents or subscripts
and the  external magnetic field and density are to be evaluated at the location of the
magnetic fibril. Note, this equation is very general; we did not assume that either of the
internal or external fields were force free, nor did we assume that the loop is torsionless.
Therefore, this equation is valid for a fibril that describes a fully 3-D curve through
an equilibrium atmosphere filled with a general 3-D external magnetic field. The term in
square brackets is the longitudinal
component and represents hydrostatic balance along field lines. The remaining terms are
transverse and correspond to the magnetic and buoyancy forces. In some ways these two
transverse terms are analogous; the last term, buoyancy, is a combination of gravity and
the net support provided by the external gas pressure, whereas the second term, the magnetic
force, is the residual Lorentz force that remains once magnetic tension and the net support
provided by the external magnetic pressure have been combined. Both forces are normal to
the surface of the tube. Our intuition may tell us that buoyancy is aligned with gravity;
but, this is only true for bodies with closed symmetric surfaces, such as a sphere. Further,
just as buoyancy can point upwards or downwards depending on whether the fibril is over- or
underdense, the net magnetic force can point up or down depending on the relative strength
of the magnetic field inside and outside the fibril.


\subsection{The Equilibrium Shape of the Fibril}
\label{subsec:EquilShape}

We now invoke the assumption that the fibril is torsionless ($\tau = 0$) and vertically
oriented, therefore lacking forces in the binormal direction. Under this assumption
we can separate the mean force equation~\eqnref{eqn:MeanForces} into only two components,

\begin{eqnarray}
\label{eqn:equil_s}
	\dpar[P]{s} &=& -g \rho \left( \unitv{z} \cdot \unitv{s} \right) \; ,
\\ \nonumber \\
\label{eqn:equil_k}
	\frac{B^2 - B_\e^2}{4\pi} R^{-1} &=& g \left(\rho - \rho_\e\right)  (\unitv{z} \cdot \unitv{k}) \; .
\end{eqnarray}

\noindent where we have used $\bvec{g}_\perp = (\bvec{g} \cdot \unitv{k}) \unitv{k}$,
which holds in equilibrium when the fibril is confined to the $x$--$z$ plane.

Equation~\eqnref{eqn:equil_s} expresses the balance of forces in the tangential
or axial direction $\unitv{s}$, whereas equation~\eqnref{eqn:equil_k} describes
the balance in the transverse direction of the principle normal $\unitv{k}$. The
axial equation is simply hydrostatic balance along magnetic field lines. The
transverse equation constrains the density contrast, $\rho-\rho_\e$, through the
balance of magnetic and buoyancy forces. This balance can be characterized by a
magnetic Bond number \citep{Jain:2012},

\begin{equation} \label{eqn:Bond}
	\varepsilon \equiv 4\pi g X \frac{\rho - \rho_\e}{B^2 - B_\e^2} \,
\end{equation}

\noindent which is a nondimensional ratio of the buoyancy to the magnetic forces.
The length scale that we have used in this definition, $2X$, is the footpoint separation,
where $x=\pm X$ is the location of each of the fibril's footpoints in the photosphere.
The magnetic Bond number appears in calculations of the Rayleigh-Taylor
instability when one or more of the fluid layers are filled with a horizontal field,
providing the critical wavenumber below which instability ensues \citep{Chandrasekhar:1961}.

Equation~\eqnref{eqn:equil_k} expresses a condition on the radius of curvature $R$ of
the fibril's axis in terms of the magnetic Bond number $\varepsilon$ and the geometry
of the fibril (i.e., the direction of the principle normal relative to gravity
$\unitv{z} \cdot \unitv{k}$),

\begin{equation} \label{eqn:ShapeEqn}
	R^{-1} = \frac{\varepsilon}{X} (\unitv{z} \cdot \unitv{k}) \; .
\end{equation}

\noindent To proceed we need expressions for the Frenet unit vectors, $\unitv{s}$ and
$\unitv{k}$, the radius of curvature, $R$, and the arclength, $s$. If $z_0(x)$ is the
height of the fibril above the photosphere, given as a function of the horizontal coordinate
$x$, and the fibril's axis is traced by the position vector $\bvec{r}(x) = x \, \unitv{x} + z_0(x) \, \unitv{z}$,
these quantities can be easily derived from equations~\eqnref{eqn:s_hat} and
\eqnref{eqn:k_hat},

\begin{eqnarray}
	\unitv{s} &=& \frac{\unitv{x} + z_0^\prime(x)\unitv{z}}{s^\prime(x)} \; ,
\\ \nonumber \\
	\label{eqn:k}
	\bvec{k}&=& -z_0^{\prime\prime}(x)  \, \frac{z_0^\prime(x) \unitv{x} - \unitv{z}}{s^\prime(x)^4} \; ,
\\ \nonumber \\
	\label{eqn:R}
	R(x) &\equiv& \left|\bvec{k}\right|^{-1} = \frac{s^\prime(x)^3}{\left|z_0^{\prime\prime}(x)\right|} \; ,
\\ \nonumber \\
	\label{eqn:sprime}
	s^\prime(x) &=& \sqrt{1 + z_0^{\prime}(x)^2} \; ,
\end{eqnarray}

\noindent where primes denote differentiation with respect to the photospheric coordinate $x$.
If we insert equations~\eqnref{eqn:k} and \eqnref{eqn:R} into equation~\eqnref{eqn:ShapeEqn},
we obtain a nonlinear ODE for the height of the fibril $z_0(x)$,

\begin{equation} \label{eqn:ODE}
	\frac{z_0^{\prime\prime}(x)}{1+z_0^\prime(x)^2} = \frac{\varepsilon}{X} \; .
\end{equation}

\noindent A quick examination of this equation reveals that the fibril will be locally
concave or convex depending on the sign of the magnetic Bond number and that points of
inflection correspond to vanishing magnetic Bond number. The magnetic Bond number is
proportional to the signed curvature. A stable fibril with a single concave arch will
have a negative Bond number everywhere, whereas a multi-arched structure will have
a magnetic Bond number that changes sign.


\subsection{Thermodynamics of the Fibril}
\label{subsec:Thermo}

The temperature, density and gas pressure within the loop can all differ from the
surrounding corona. We define the contrasts is these properties in the following
manner:

\begin{eqnarray} \label{eqn:Def_rho_con}
	\Delta\rho(s) &\equiv& \rho(s)-\rho_\e(s) \; ,
\\ \nonumber \\
	\Delta P(s) &\equiv& P(s)-P_\e(s) \; ,
\\ \nonumber \\
	\label{eqn:Def_T_con}
	\Delta T(s) &\equiv& T(s)-T_\e(s) \; .
\end{eqnarray}

\noindent Technically the external variables are all functions of two spatial coordinates---e.g.,
$x$ and $z$. In all of these definitions, however, the external variable is evaluated at the
location of the fibril. We indicate this by expressing these variables as a function of
the pathlength alone. Only the contrasts in the mass density $\Delta \rho$, gas pressure
$\Delta P$, and magnetic pressure $\Delta B^2/8\pi$ play an active role in the force balance.
The temperature contrast $\Delta T$ is a dependent variable that can be derived from these
active variables post facto. There are four primary equations that provide all of the inter-relations
between these properties of the loop,

\begin{eqnarray}
	\label{eqn:4E-Pcont}
	{\textstyle\rm Pressure~Continuity} \qquad \D \Delta P &=& -\frac{\Delta B^2}{8\pi}  \; ,
\\ \nonumber \\
	\label{eqn:4E-Hydrostatics}
	{\textstyle\rm Hydrostatic~Balance} \quad\; \D \der[\Delta P]{z_0} &=& -g \Delta\rho  \; ,
\\ \nonumber \\
	\label{eqn:4E-Transverse}
	{\textstyle\rm Transverse~Force~Balance} \qquad\quad \D \frac{\varepsilon}{X} &=& -\frac{g}{2} \, \frac{\Delta\rho}{\Delta P} = \frac{z_0^{\prime\prime}}{1+(z_0^{\prime})^2}  \; ,
\\ \nonumber \\
	\label{eqn:4E-FlareRate}
	{\textstyle\rm Field~Strength~Deviation} \quad\; \D \frac{\Delta B^2}{8\pi} &=& \alpha \frac{B_\e^2}{8\pi}  \; .
\end{eqnarray}

The first of these is a direct consequence of equation~\eqnref{eqn:PressureCont}. The second
results from transforming equation~\eqnref{eqn:equil_s} from the pathlength variable $s$ to the
fibril height $z_0$. The third relation, equation~\eqnref{eqn:4E-Transverse}, arises from a
combination of the definition of the magnetic Bond number~\eqnref{eqn:Bond} and the equation
of transverse force balance~\eqnref{eqn:ODE}, where the magnetic-pressure contrast has been
replaced through use of pressure continuity, equation~\eqnref{eqn:4E-Pcont}. The last equation
is a restatement of equation~\eqnref{eqn:MatchedFlare} which ensures that loop and external
magnetic field share a common bounding flux surface. The temperature contrast $\Delta T$ can
be found by using the ideal gas law ($P = R_{\rm gas}\rho T$),

\begin{equation} \label{eqn:DeltaT}
	\Delta T = \frac{g(h-H_\e)}{R_{\rm gas}} \frac{\Delta\rho}{\rho} \; ,
\end{equation}

\noindent where $H_\e$ is the external corona's pressure scale height, $H_\e = R_{\rm gas} T_\e /g$,
and $h$ is the scale height of the gas-pressure contrast,

\begin{equation} \label{eqn:Def_h}
	h^{-1} \equiv -\frac{2\varepsilon}{X} = - \frac{1}{\Delta P}\der[\Delta P]{z_0} \; .
\end{equation}

Conveniently, these equations can be solved analytically to obtain the contrast variables
as a function of the fibril's shape. If we combine equations~\eqnref{eqn:4E-Hydrostatics}
and \eqnref{eqn:4E-Transverse}, we obtain a differential equation that relates the
pressure contrast to the derivative of the fibril's height,

\begin{equation} \label{eqn:dlnPdz0}
	\frac{1}{\Delta P} \der[\Delta P]{z_0} = \frac{2 z_0^{\prime\prime}(x)}{1+z_0^{\prime}(x)^2} \; .
\end{equation}

\noindent This equation can be directly integrated to obtain,

\begin{equation} \label{eqn:DeltaP}
	\Delta P = -\alpha C \left[1+(z_0^{\prime})^2\right] \; ,
\end{equation}

\noindent where $C$ is a positive integration constant that can be fixed in a variety
of ways. One could apply a photospheric boundary condition or one could use spectroscopic
observations to fix the temperature or density contrast at the apex of the loop. Strictly
speaking, the integration constant is the product $\alpha C$. We have included the
field-strength deviation in this product for later convenience and in order to remind the reader
that the thermodynamic contrasts arise from a small deviation from the force-free equilibrium.
If we subsequently use the equations of pressure continuity~\eqnref{eqn:4E-Pcont} and
hydrostatic balance~\eqnref{eqn:4E-Hydrostatics}, we can derive equations for the mass
density and magnetic-pressure contrasts,

\begin{eqnarray}
	\label{eqn:DeltaRho}
	\Delta\rho &=& \frac{2\alpha C}{g} z_0^{\prime\prime} \; ,
\\
	\frac{\Delta B^2}{8\pi} &=& \alpha C \left[1+(z_0^{\prime})^2\right] \; .
\end{eqnarray}

\noindent Thus, if we are provided the shape of the loop, we can derive all of the
loop's contrast variables. Different values of the integration constant
$\alpha C$ result in different temperature profiles along the loop. Equation~\eqnref{eqn:DeltaT}
can be rewritten to show how the temperature contrast depends on the geometry of
the loop and the constant $\alpha C$,

\begin{equation} \label{eqn:DeltaT_alt}
	\Delta T = - \alpha C T_\e \, \frac{\left[1 + (z_0^\prime)^2 + 2 H_\e z_0^{\prime\prime}\right]}{P_\e + 2 \alpha C H_\e z_0^{\prime\prime}}  \; .
\end{equation}

\noindent Note that the integration constant $C$ was chosen such that it appears
only in the combination of $\alpha C$ in the equations above. Thus, one needs
only a single constraint to define the thermodynamic variables.


\subsection{Self-Consistent External Magnetic Field}
\label{subsec:Bext}

Given the pressure contrast $\Delta P$, we can use equations~\eqnref{eqn:4E-Pcont},
\eqnref{eqn:4E-FlareRate}, and \eqnref{eqn:DeltaP} to express the external field
strength at the location of the fibril in terms of the geometry of the fibril,

\begin{equation} \label{eqn:Bext}
	\frac{B_\e^2(s)}{8\pi}  = C \left[1+(z_0^{\prime})^2\right] \; .
\end{equation}

\noindent This last equation shows that the integration constant $C$ is really just
the value of the external magnetic pressure at the apex of the loop (hence $C$ is
positive). Further, the equation demonstrates that not all external fields are
self-consistent with an embedded fibril. If we select an arbitrary field line within
a general magnetic field, equation~\eqnref{eqn:Bext} may not be satisfied for a constant $C$.
This becomes readily apparent when we express the direction of the tangent vector in
terms of the angle $\theta$ that it makes with the $x$-axis,
$\unitv{s} \equiv \cos\theta \, \unitv{x} + \sin\theta\, \unitv{z}$. With a little
basic manipulation, we find the following relation between the angle $\theta$ and
the height function $z_0$:

\begin{equation}
	\sec^2\theta = 1 + (z_0^\prime)^2 \; .
\end{equation}

\noindent Subsequently, one can easily show that equation~\eqnref{eqn:Bext} is equivalent
to the statement that the $x$-component of the external magnetic field is constant along
the length of the fibril,

\begin{equation}
	\unitv{x}\cdot\bvec{B}_\e = B_\e \cos\theta = \sqrt{8\pi C} \; .
\end{equation}

\noindent Of course, not all force-free fields will have this property.

If needed, a self-consistent external magnetic field can be derived by noting that the shape of the
fibril and the strength of the magnetic field at the location of the fibril provide a boundary
condition for the coronal magnetic field that occupies the bulk of the domain. For simplicity
we assume that the external magnetic field is potential and 2D. Therefore, since the field
is solenoidal and 2D it can be described by a flux function $\Psi_\e$,

\begin{equation}
	\bvec{B}_\e = \grad\times\left(\Psi_\e \unitv{y}\right)  = -\dpar[\Psi_\e]{z} \, \unitv{x} + \dpar[\Psi_\e]{x} \, \unitv{z} \; .
\end{equation}

\noindent Since the field is potential the flux function must be harmonic,

\begin{equation} \label{eqn:Harmonic}
	\nabla^2\Psi_\e = 0 \; .
\end{equation}

\noindent Therefore, we can find a self-consistent external field solution by solving
equation~\eqnref{eqn:Harmonic} under the boundary condition

\begin{displaymath} \label{eqn:BC}
	\bvec{\nabla}\Psi_\e = B_\e(s) \, \unitv{k}(s) \; ,
\end{displaymath}

\noindent applied at the location of the fibril with $B_\e(s)$ fixed by equation~\eqnref{eqn:Bext}.
This ensures that the external field is both parallel to the fibril and the field strength
is given by $B_\e(s)$. There is an infinity of such solutions, all with different photospheric
boundary conditions.


\section{A Model with Uniform Magnetic Bond Number}
\label{sec:ConstantEpsilon}

For our first example solution we wish to consider one that is both simple and fully
analytic. Therefore, we consider a fibril which has constant magnetic Bond number
$\varepsilon$ along its length. For constant $\varepsilon$, the transverse force
equation~\eqnref{eqn:4E-Transverse} can be integrated twice to obtain the height
of the fibril $z_0(x)$. Similarly, using the result of the first of these integrations,
equation~\eqnref{eqn:sprime} can be integrated to obtain the arclength $s(x)$. We
choose the three constants of integration such that the footpoints occur at $x = \pm X$,
the fibril is symmetric about $x=0$, and the footpoint at $x=-X$ corresponds to $s=0$,

\begin{equation}
	z_0(\pm X) = z_0^\prime(0) = s(-X) = 0 \; .
\end{equation}

\noindent With these conditions, we find the following solution for the geometrical
properties of the fibril:
 
\begin{eqnarray}
	\label{eqn:z0}
	z_0(x) &=& -\frac{X}{\varepsilon} \ln\left(\frac{\cos\left(\varepsilon x/X\right)}{\cos \varepsilon}\right) \; ,
\\ \nonumber \\
	\label{eqn:dz0dx}
	z_0^\prime(x) &=& \tan\left(\varepsilon x/X\right) \; ,
\\ \nonumber \\
	\label{eqn:R0}
	R(x) &=& -\frac{X}{\varepsilon} \sec\left(\varepsilon x/X\right) \; ,
\\ \nonumber \\
	\label{eqn:arclength}
	s(x) &=& \frac{X}{2\varepsilon} \ln\left( \frac{1+\sin\varepsilon}{1-\sin\varepsilon} \; \;
			\frac{1+\sin\left(\varepsilon x/X\right)}{1-\sin\left(\varepsilon x/X\right)} \right) \; .
\end{eqnarray}

\noindent Figure~\ref{fig:ConstBond} displays solutions for several different values
of $\varepsilon$. It is clear that fibril posesses a nonconstant radius of curvature,
with strong curvature near the apex and weaker curvature in the legs.

The length of the fibril $L$ is obtained by inserting the rightmost footpoint position
$x = X$ into equation~\eqnref{eqn:arclength},

\begin{equation} \label{eqn:Length}
	L = \frac{X}{\varepsilon} \ln\left( \frac{1+\sin\varepsilon}{1-\sin\varepsilon} \right) \; .
\end{equation}

\noindent Two interesting limits of this equation exist. As $\varepsilon \to 0$ the
length of the loop converges to the footpoint separation $L \to 2X$. This arises
because the loop becomes straight, flat, and confined to the photosphere.  As
$\varepsilon \to \pm \pi/2$ the loop length diverges logarithmically because the height
of the loop grows without bound. This can be seen by recognizing that the loop reaches
its apex at its center ($x=0$), therefore achieving a maximum height of

\begin{equation} \label{eqn:zapex}
	z_{\rm apex} = z_0(0) = \frac{X}{\varepsilon} \ln\left(\cos\varepsilon\right) \; .
\end{equation}

\noindent One can easily see that $\varepsilon = \pm\pi/2$ corresponds to a logarithmic
singularity in the height. Clearly the height $z_0(x)$ should be a positive function
for the range $x \in (-X,~X)$. With a little thought, from equation~\eqnref{eqn:z0} we
can see that this is only possible if $\varepsilon \in (-\pi/2, 0)$, once again emphasizing
that a fibril with a single concave arch has a negative magnetic Bond number.


\subsection{Thermodynamics of a Fibril with Uniform Magnetic Bond Number}
\label{subsec:Thermodynamics_ConstEps}

The contrast variables have a common geometrical factor, $1+(z_0^\prime)^2$, appearing
in their functional forms. For the model with uniform magnetic Bond number this factor
takes on a simple form. Using equations~\eqnref{eqn:z0} and \eqnref{eqn:dz0dx}, we find

\begin{equation}
	1 + (z_0^\prime)^2 = \sec^2\left(\varepsilon x/X\right) = \sec^2\varepsilon \; \exp\left(-z_0/h\right) \; .
\end{equation}

\noindent Using this expression and the equations derived in subsection~\S\ref{subsec:Thermo}
we can solve for all of the contrast variables and the external magnetic pressure in
terms of the scale height $h$ that depends on the Bond number, $h=-X/2\varepsilon$,

\begin{eqnarray}
	\label{eqn:ConstEps_DP}
	\Delta P &=& -\frac{\Delta B^2}{8\pi} = -\alpha C \, \exp\left(\frac{z_{\rm apex}-z_0}{h}\right) \; ,
\\ \nonumber \\
	\label{eqn:ConstEps_Drho}
	\Delta\rho &=& -\frac{\alpha C}{gh} \, \exp\left(\frac{z_{\rm apex}-z_0}{h}\right) \; ,
\\ \nonumber \\
	\label{eqn:ConstEps_DT}
	\Delta T &=& T_\e \left(\frac{h}{H_\e} - 1\right) \frac{\Delta\rho}{\rho_\e+\Delta\rho_\e} \; ,
\\ \nonumber \\
	\label{eqn:ConstEps_Be}
	\frac{B_\e^2}{8\pi} &=& C \, \exp\left(\frac{z_{\rm apex}-z_0}{h}\right) \; .
\end{eqnarray}

\noindent The density and pressure contrasts, as well as the external magnetic
pressure, are revealed to be exponential functions of height, all with the same constant
scale height $h$. The density and temperature contrast are illustrated in Figure 3,
for an isothermal exterior with a scale height of $H_\e = 75$ Mm. The constant $\alpha C$ has
been chosen such that the fractional density contrast at the apex of the loop is
$\Delta\rho/\rho_\e = \pm 0.01$. The solid curves correspond to overdense loops
and the dotted curves to underdense loops, with the different colors corresponding
to loops with different magnetic Bond numbers. We will see in the discussion section
\S\ref{sec:Discussion} that overdense loops have been heated relative to the surrounding
corona and underdense loops have been cooled. Therefore, overdense loops are probably
more physically relevant.

For an isothermal exterior, where $H_\e$ is a constant, one could choose the magnetic
Bond number such that $h = H_\e$. For such a model, the temperature contrast would be
identically zero all along the length of the loop and the plasma parameter $\beta = 8\pi P/B^2$
would be a constant both inside and outside the loop, while the density contrast
remains nonzero. The condition $H_\e = h$ marks a transition between disparate
behavior. Loops with $H_\e < h$ have fractional density and temperature contrasts that
decrease with height, whereas loops with $H_\e > h$ have fractional density and temperature
contrasts that increase with height.


\subsection{A Self-Consistent External Magnetic Field}
\label{subsec:ExtField}

In the previous subsection we discovered that a fibril with constant magnetic Bond
number requires that the external magnetic pressure have a constant scale height as
we slide along the fibril. This suggests that we seek a solution in which the
magnetic pressure is only a function of height and varies exponentially. A simple
harmonic solution that satisfies this requirement is a sinusoid in the horizontal
$x$-direction multiplied by a decaying exponential in height $z$,

\begin{equation} \label{eqn:Psi}
	\Psi_\e(x,z) = \kappa^{-1} \, \tilde{B}_\e  \cos(\kappa x) \, e^{-\kappa z} \; .
\end{equation}

\noindent In this solution, $2\pi/\kappa$ is the horizontal repetition length of
the field ($\Psi_\e$ changes sign every $\pi/\kappa$). With such a flux function,
the magnetic field is as follows:

\begin{equation}
	\bvec{B}_\e(x,z) = \tilde{B}_\e \left[\cos(\kappa x) \unitv{x} - \sin(\kappa x) \unitv{z} \right] \, e^{-\kappa z} \; .
\end{equation}
 
\noindent Since the magnetic field is periodic in the $x$-direction, we will only
consider the arcade that is symmetric about the origin and exists in the range
$x \in (-D, D)$, where $D = \pi/(2\kappa)$.  The magnetic pressure for such a
field depends only on height and falls off exponentially with a scale height of
$(2\kappa)^{-1}$,

\begin{equation}
	\frac{B_\e^2}{8\pi} = \frac{\tilde{B}_\e^2}{8\pi} \, e^{-2\kappa z} \; .
\end{equation}

\noindent In order to satisfy equation~\eqnref{eqn:ConstEps_Be}, we must insist
that the external field's horizontal wavenumber is proportional to the magnetic
Bond number and that the photospheric value of the field strength depends on $C$,

\begin{eqnarray}
	\label{eqn:Be_wavenumber}
	\kappa &=& -\frac{\varepsilon}{X} = \frac{2}{h} \; ,
\\ \nonumber \\
	\frac{\tilde{B}_\e}{8\pi} &=& C \exp\left(\frac{z_{\rm apex}}{h}\right) \; .
\end{eqnarray}

Lines of constant $\Psi_\e$ in the $x$--$z$ plane denote field lines. The height above the
photosphere, $z_\e$, of a field line with a specified value of the flux function is
easily obtained by inverting equation~\eqnref{eqn:Psi},

\begin{equation} \label{eqn:ze}
	z_\e(x; \Psi_\e) = \kappa^{-1} \ln \left( \frac{\cos(\kappa x)}{\cos(\kappa x_\Psi)}\right) \; , 
\end{equation}

\noindent where the footpoints intersect the photosphere at $x = \pm x_\Psi$,

\begin{equation} \label{eqn:footpoints}
	x_\Psi \equiv \kappa^{-1} \cos^{-1}\left(\frac{\kappa\Psi_\e}{\tilde{B}_\e}\right) \; .
\end{equation}

\noindent We can clearly see from equation~\eqnref{eqn:z0} that this potential field solution
for the external corona is consistent with the loop model, equation, if the
magnetic Bond number is inversely proportional to the repetition length of the external
magnetic field, $\varepsilon = -\kappa X$, which is a condition we have already imposed
to ensure that the external field strength has the appropriate value along the fibril.

The field lines of this potential field are illustrated in Figure~\ref{fig:ConstBondField}.
In the beginning of this section we discovered that stability requires that
the magnetic Bond number be bounded $\varepsilon \in (-\pi/2, 0)$. In the context of
this specific model, we can see that the magnetic Bond number selects the field line
within the coronal field that corresponds to the axis of the fibril. Fibrils with different
footpoint separations and magnetic Bond numbers are shown as various colored field
lines in Figure~\ref{fig:ConstBondField}. A small value of the magnetic Bond number corresponds
to a short, flat, low-lying loop confined very near the origin, while a magnetic Bond
number near the lower limit is a tall, long loop, with foot-points approaching
the outer edge of the arcade. In fact, in the limit $\varepsilon \to -\pi/2$, the loop
is infinitely tall with vertical legs located at $x = \pm X = \pm D$. Therefore, there
is a direct mapping between the magnetic Bond number and the value of the flux function
that corresponds to the loop's axis,

\begin{equation}
	\Psi_{\rm fibril} = (2D/\pi)\, \tilde{B}_\e  \cos\varepsilon = h \sqrt{\frac{\pi C}{2}} \; ,
\end{equation}

\noindent and one can choose to place the loop along any field line in the coronal
model by dialing the magnetic Bond number between $-\pi/2$ and 0.


\section{A Model with Uniform Radius of Curvature}
\label{sec:ConstantRadius}

Another simple model to consider is a fibril with a constant radius of curvature.
This model is of course an example of a fibril with a nonuniform magnetic Bond
number. However, it is still a single concave arch and therefore the Bond number
will be negative everywhere. With a constant radius of curvature $R$,
equation~\eqnref{eqn:ShapeEqn} simplifies to

\begin{equation}
	\varepsilon = -\frac{X}{z_0 - z_c} \; ,
\end{equation}

\noindent where $z_c$ is the height above the photosphere of the fibril's center
of curvature. From this expression we can see that the magnetic Bond number diverges
when $z_0 = z_c$ resulting in a divergence of all the contrast variables at the
same location. Therefore, for a physically meaningful solution we must insist that
the center of curvature lies below the photosphere, i.e., $z_c < 0$. The fibril is
therefore a circular arc that is less than a full semi-circle and intersects the
photosphere at oblique angles.

Using the procedure outlined in subsection~\S\ref{subsec:Thermo} we derive
all of the contrast variables and the external magnetic pressure from the magnetic
Bond number,

\begin{eqnarray}
	\Delta P &=& -\alpha C \left(\frac{R}{z_0-z_c}\right)^2 \; ,
\\ \nonumber \\
	\Delta\rho &=& -\frac{2\alpha C}{gR} \, \left(\frac{R}{z_0-z_c}\right)^3 \; ,
\\ \nonumber \\
	\label{eqn:ConstR_DT}
	\Delta T &=& T_\e \, \left(\frac{z_0-z_c}{2H_\e} - 1\right) \frac{\Delta\rho}{\rho_\e + \Delta\rho} \; ,
\\ \nonumber \\
	\label{eqn:ConstR_Be}
	\frac{B_\e^2}{8\pi} &=& C \, \left(\frac{R}{z_0-z_c}\right)^2 \; .
\end{eqnarray}

\noindent Figure~\ref{fig:TandD_circ} illustrates the density and temperature contrasts
for loops with radii of curvature spanning 25--125 Mm. Each is embedded in an isothermal
corona with a scale height of 75 Mm. The integration constant $\alpha C$ has been chosen
such that the fractional density contrast is $\Delta\rho/\rho_\e = \pm 0.01$ at the apex
of the loop. For all of the loops the magnitude of the density contrast
decreases with height. For the parameters chosen for the figure, the overdense loops
tend to be cold compared to their surroundings, although those that are sufficiently
large ($R > 2H_\e$) can have a warm crest. Similarly, underdense loops tend to be hot
with the tallest and widest loops possessing a cool crest. If a change of sign occurs,
it will happen at a distance of two scale heights above the center of curvature.

Many external field solutions could be found that possess circular field lines;
however, most will not possess the needed functional form for the external field
strength along the loop. Consider the field generated by a line current located
at the center of curvature. Such a line current generates a field with concentric
field lines that have constant field stength as one moves along a line. We can see that
this field fails because equation~\eqnref{eqn:ConstR_Be} dictates that the field
strength must decrease with height along the fibril. The functional form for $B_\e$
given by that equation and the fact that the fibril is a circular arc, suggests
that we seek a solution in polar coordinates with the following form,

\begin{eqnarray}
	\Psi_\e(r,\phi) &=& \tilde{B}_\e \, \frac{z_c}{R} \, \frac{r-R}{\sin\phi} \; ,
\\
 \nonumber \\
	\bvec{B}_\e &=& \frac{1}{r}\dpar[\Psi_\e]{\phi} \, \unitv{r} - \dpar[\Psi_\e]{r} \, \unitv{\phi}
\\ \nonumber \\ \nonumber
	&=& -\frac{z_c}{R} \, \frac{\tilde{B}_\e}{\sin^2\phi} \left( \frac{r-R}{r} \, \cos\phi \, \unitv{r} + \sin\phi \, \unitv{\phi}  \right) \; ,
\end{eqnarray}

\noindent where $r^2 = x^2 + (z-z_c)^2$ and $\sin\phi = (z-z_c)/r$. The origin
of the polar coordinate system is located at the center of curvature, $R$ is
the radius of curvature of the fibril, and $\phi = 0$ points in the positive
$x$-direction. Once again, $\tilde{B}_\e$ is the value of the external magnetic
field strength at the footpoints. We can verify that this particular potential
field has a circular field line at $r=R$ by noting that the flux function
is constant there (in fact, $\Psi_\e = 0$ on the fibril). Direct evaluation
further verifies that the magnetic pressure when evaluated at the location of
the fibril has the requisite functional form, equation~\eqnref{eqn:ConstR_Be},
as long as

\begin{equation}
	\frac{\tilde{B}_\e^2}{8\pi} = \frac{CR^2}{z_c^2} \; .
\end{equation}

The field lines for this potential field are illustrated in Figure~\ref{fig:ConstCurvField}.
The external field lines are drawn in black and the
circular fibril is marked in blue. Note, unlike the constant magnetic Bond number
model presented in section~\S\ref{sec:ConstantEpsilon} where one could dial the
control parameter $\varepsilon$ to change where the fibril appears in the corona,
only one field line in Figure~\ref{fig:ConstCurvField} meets the prerequisite
of constant curvature. Thus, while self-similar, this external magnetic-field
model differs for fibrils with different radii of curvature.

\section{Discussion}
\label{sec:Discussion}

Our equilibrium model is predicated on the implicit assumption that force balance
is achieved at a much faster time scale than radiative cooling and thermal conduction
redistribute heat. Hydrostatic equilibrium should be established on a dynamical
time scale in between the free-fall time and the acoustic crossing time. Thus, for
a loop that reaches 50 Mm above the photosphere and for a pressure scale height of
80 Mm, the dynamical time scale lies between 10--40 minutes. Transverse force balance
is achieved by mass redistribution along the loop and should therefore be established
on a similar time scale. Estimates of the radiative cooling and conduction times vary,
but for active region loops radiative times can be on the order of 1--40 hr and conduction
times are typically 1 hr \citep{Aschwanden:2005}. Therefore, we expect that immediately after
an impulsive heating event, a coronal loop will quickly re-establish force balance
and will stay in force balance as the loop slowly cools. In the following section
we discuss the implications of this balance.


\subsection{Role of the Magnetic Bond Number}
\label{subsec:MagBondNum}

The shape of a coronal loop is determined by the force-free equilibrium of the
surrounding corona. Given this shape, the mass along the loop must redistribute
itself such that the buoyancy forces exactly oppose the Lorentz forces acting on
the loop. The resulting balance of forces can be succinctly characterized by the 
magnetic Bond number which is fully specified by the height of the loop above the
photosphere as a function of position. In equilibrium, the magnetic Bond number
is proportional to the signed curvature of the loop, and therefore concave curvature
results in a negative magnetic Bond number and convex curvature in a positive Bond
number. Of course, inflection points correspond to locations where the Bond number
vanishes. Therefore, a loop comprised of a single concave arch must have negative magnetic
Bond number everywhere if in equilibrium. Equilibrium loops that possess multiple
arches will have negative Bond number in the crests and positive Bond number in the
dips or troughs.

All loops with the same geometric shape have exactly the same profile of the magnetic
Bond number. However, these loops can be achieved in a continuum of different ways, all
varying in the distribution of mass and field strength. This continuum is represented
by different values of an integration constant $\alpha C$ which is a direct measure
of the magnetic-pressure contrast at the apex of the loop. For example, an underdense
loop with enhanced magnetic-field strength can have exactly the same Bond number
as an overdense loop with reduced field strength. While the shape of these two
loops will be the same, the thermodynamic and magnetic properties of these two
loop might be very different.


\subsection{Mass Redistribution along the Loop}
\label{subsec:MassRedist}

For a loop with enhanced field strength, the mass density is proportional 
to the second derivative of the height function with respect to the horizontal
photospheric coordinate, equation~\eqnref{eqn:DeltaRho}. Thus, we can immediately
deduce that peaks or crests in a loop will have a mass deficit while dips or troughs
have mass accumulation. Furthermore, if we define the total mass deviation between
two points as the integrated density contrast,

\begin{equation}
	\Delta M(s_1,s_2) \equiv \int_{s_1}^{s_2} \Delta\rho(s) \, A(s) \, ds \; ,
\end{equation}

\noindent then the mass deviation vanishes if we choose two points that correspond
to extrema in height. This can be revealed by noting that flux conservation within
the loop requires the following:

\begin{equation}
	A(s) = \frac{A_{\rm apex}}{\left[1+(z_0)^2\right]^{1/2}} \; ,
\end{equation}

\noindent where $A_{\rm apex}$ is the cross-sectional area of the loop at the apex.
If we change variable from the pathlength to the photospheric $x$ coordinate and
replace the density contrast using equation~\eqnref{eqn:DeltaRho} the
integral can be evaluated trivially,

\begin{equation}
	\Delta M(x_1,x_2) = \frac{2\alpha C A_{\rm apex}}{g} \, \left[z_0^\prime(x_2) - z_0^\prime(x_1)\right] \; ,
\end{equation}

\noindent Thus, the net mass deviation between a crest and a neighboring
trough is zero and the crest drains mass into the dips. For a loop comprised
of a single concave arch, the mass deviation between the two footpoints is negative. Mass must
drain out of the loop through the photosphere into the solar interior. In the opposite
case, a loop with a reduced field strength, the mass density is proportional
to the negative of the second derivative. The peaks in such a loop must have enhanced
density so that buoyancy can oppose the upward Lorentz force. This obviously implies
that the gas pressure must increase throughout the loop to increase the hydrostatic
support. The additional mass is lifted from the photosphere into the loop.


\subsection{Temperature Profile}
\label{subsec:LoopTemp}

Observations of the temperature along the length of a coronal loops have suggested
that some loops might be nearly isothermal \citep{Aschwanden:2001}. For the relatively
small density contrasts that have been explored here, the temperature contrast is
also relatively small and the temperature is perforce roughly isothermal as long
as the external corona is isothermal. However, we wish to point out that the temperature
contrast itself is markedly constant over height for a signficant fraction of the models
we have illustrated. Three of the models with constant magnetic Bond number, appearing
in Figure 3, have a temperature variation of less than a 30\% percent from footpoint to
apex. These three correspond to the models with the squattest loops.

Given a specific shape for a coronal loop there exists a family of possible solutions,
each characterized by a different value of $\alpha C$, the integration constant---see
equations~\eqnref{eqn:DeltaP}--\eqnref{eqn:DeltaT_alt}. When illustrating our models,
we chose to fix this parameter by specifying the fractional density contrast at the apex
of the loop. We wish to point out, however, that the parameter $\alpha C$ is really a
measure of the total heat that has been input into the coronal loop. The excess heat
contained by the loop per unit length is given by

\begin{equation}
	E(s) = c_v A(s) \left[\rho(s)T(s) - \rho_\e(s) T_\e(s)\right] = (\gamma-1)^{-1} A(s) \Delta P(s)  \; ,
\end{equation}

\noindent where $c_v$ is the specific heat at constant volume and $\gamma$ is the adiabatic
exponent. We can obtain the total
excess heat contained by the loop by integrating $E(s)$ over the loop's length. If we
eliminate the gas-pressure contrast by using equation~\eqnref{eqn:DeltaP}, we reduce
the integral to a constant factor times a positive definite integral that depends only
on the geometry of the loop,
 
\begin{equation}
	\int_0^L E(s) \, ds = -\frac{\alpha C A_{\rm apex}}{\gamma-1}  \,
		\int_{-X}^X \left[ 1+\left(z_0^\prime\right)^2\right] \, dx \; ,
\end{equation}

\noindent Since the constant $C$ and the integral appearing on the right hand side of the previous
equation are always positive, we can immediately discern two facts: (1) Heated 
loops correspond to those with a negative field-strength deviation $\alpha$ and cooled loops
to those with a positive value. Thus, heated loops are those that are undermagnetized and
overdense and cooled loops are overmagnetized and underdense. (2) The integration constant
$C$ is a measure of the magnitude of the heat that has been injected into or extracted from the
loop. Obviously, heat deposition is more likely to be physically relevant. The
family of solutions defined by different values of $C$ form a continuum of loop models with
different amounts of stored heat.  For example, after a large impulsive heating event, the heat
content of the loop jumps but force balance is quickly re-established. This
requires a redistribution of mass (and heat) along the loop and the resulting balance is
characterized by a nonzero value of $C$. Then as radiative cooling and conduction slowly
dissipate the loop's excess heat, the loop responds by moving through a sequence of equilibria
while simultaneously maintaining force balance. This sequence is represented by slow
temporal attenuation of the parameter $C$.


\subsection{Implications for Loop Seismology}
\label{subsec:LoopOsc}

Seismology of a coronal loop is only sensitive to the distribution of wave speed
along the loop \citep[e.g., ][]{Jain:2012}. If we assume that the observed waves
are kink oscillations and that they propagate at the kink speed,

\begin{equation}
	c_K^2 = \frac{B^2 + B_\e^2}{4\pi(\rho + \rho_\e)} \; ,
\end{equation}

\noindent then, at best, a seismic analysis can provide information about the ratio
of the field strength to the density. It is impossible from only a seismic analysis
to measure the field strength profile independently of the density profile. However,
with additional constraints, provided either from observations or from basic physical
principles, one might be able to disentangle field-strength and density variations
along the loop. The force-balance model constructed here provides just such a constraint.
The density contrast and magnetic-pressure contrast are related to the geometry of the
loop through the magnetic Bond number,

\begin{equation}
	\frac{\Delta B^2}{4\pi} = \frac{gX}{\varepsilon} \Delta\rho = gX \frac{1+(z_0^\prime)^2}{z_0^{\prime\prime}} \Delta\rho \; .
\end{equation}

If we assume that the interior magnetic field is nearly the same as the external
field  (hence the field-strength deviation is small), we can use this previous equation
to express the kink speed solely in terms of the density contrast, the exterior density
and the loop's geometry,

\begin{equation}
	c_K^2 = 2C \frac{1 + (z_0^\prime)^2}{\rho_\e + (\alpha C/g) z_0^{\prime\prime} } \; .
\end{equation}

\noindent If we further assume that the exterior corona is isothermal, $\rho_\e = \tilde{\rho}_\e \exp(-z/H_\e)$,
this expression has only four free parameters: the integration constant $C$, the field-strength
deviation $\alpha$, the coronal scale height $H_\e$ and photospheric value of the exterior
density $\tilde{\rho}_\e$. Thus, even the measurement of only a few mode frequencies
that provide different spatial averages of the kink speed should be sufficient to either
fix these parameters or to verify a choice made using other information.


\subsection{Conclusions}
\label{subsec:Conclusions}

We have developed a model of curved coronal loops that self-consistently couples
deviations from the force-free state to the thermodynamic properties. The coupling
is accomplished by requiring that the loop is in equilibrium and the Lorentz force
arising from the deviation from a force-free field is balanced by buoyancy. This
links the curvature of the loop to the loop's mass-density contrast and hence to
the pressure and temperature through hydrostatic balance and the ideal gas law.
The end result is that specification of the loop's geometry is sufficient to derive
the temperature and density profiles to within a single integration constant.
This integration constant can be selected in a variety of ways, either seismically
through estimates of the mean kink-wave speed or spectroscopically through estimates
of the density or temperature contrast at one position along the loop. Further,
this integration constant is a direct assessment of the thermal history of the loop
providing an estimate of the excess heat contained by the loop relative to the
surrounding corona.


\acknowledgements

This work is supported by NASA, MSRC (University of Sheffield) and STFC (UK). BWH
acknowledges NASA grants NNX08AJ08G, NNX08AQ28G, and NNX09AB04G.




\begin{figure*}%
        \epsscale{0.5}%
        \plotone{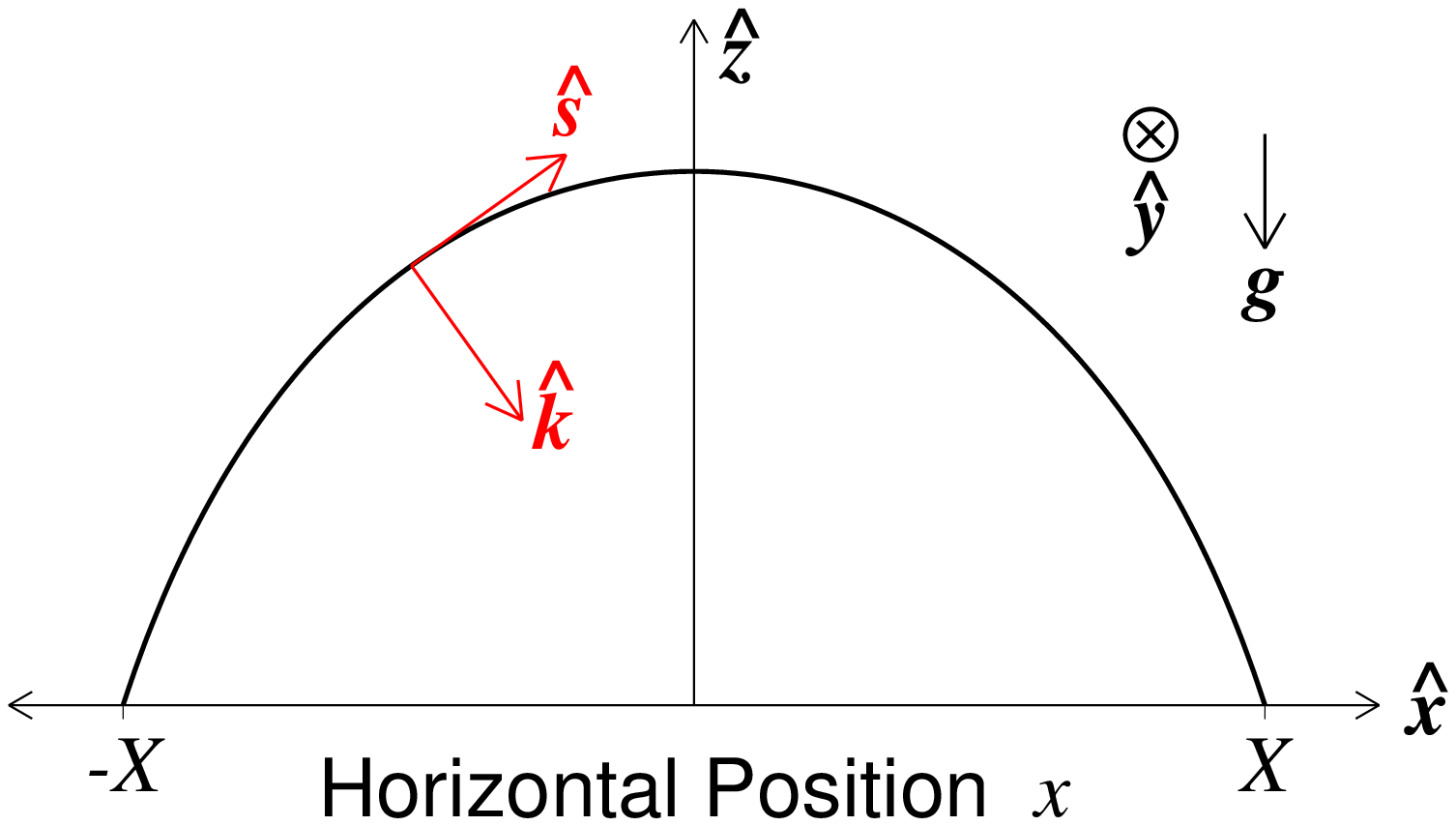}%
        \caption{\small The geometry of the coronal loop, with the loop's axis shown
in black. The photosphere corresponds to the $x$--$y$ plane, while the loop is confined
to the $x$--$z$ plane. The $y$-axis points into the page. The local Frenet coordinates
are indicated in red. The local tangent vector is $\unitv{s}$ and the principle
normal $\unitv{k}$ lies in the direction of curvature. The binormal is everywhere
constant and pointed in the $\unitv{y}$ direction.
\label{fig:Geometry}}%

\end{figure*}%


\begin{figure*}%
        \epsscale{1.0}%
        \plotone{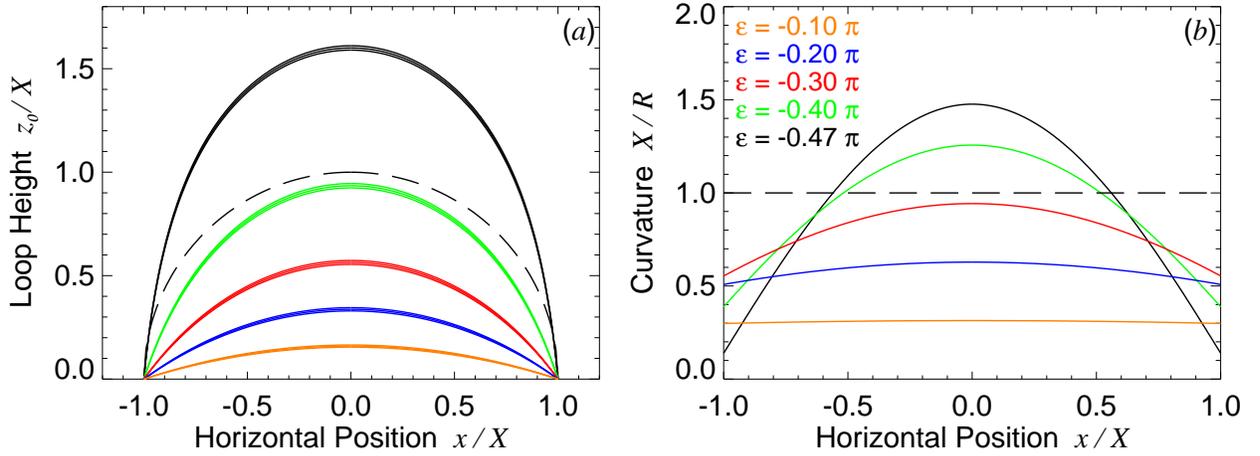}%
        \caption{\small Equilibrium ($a$) height and ($b$) curvature of coronal loops
with differing ratios of buoyancy to magnetic forces, i.e., differing values of
the magnetic Bond number $\varepsilon$. The value of $\varepsilon$ associated with each color
is indicated in the right panel.  In both panels the dashed line corresponds to a
semi-circle with constant radius of curvature $R = X$. Loops with weak buoyancy
($\left|\varepsilon\right|<< 1$) are flat with large radius of curvature everywhere. Loops
with substantial buoyancy ($\varepsilon \approx -\pi/2$) are tall with small curvature in
the legs and large curvature at its apex. Loops with positive $\varepsilon$ or with
$\varepsilon < -\pi/2$ are unstable.
\label{fig:ConstBond}}%

\end{figure*}%


\begin{figure*}%
        \epsscale{1.0}%
        \plotone{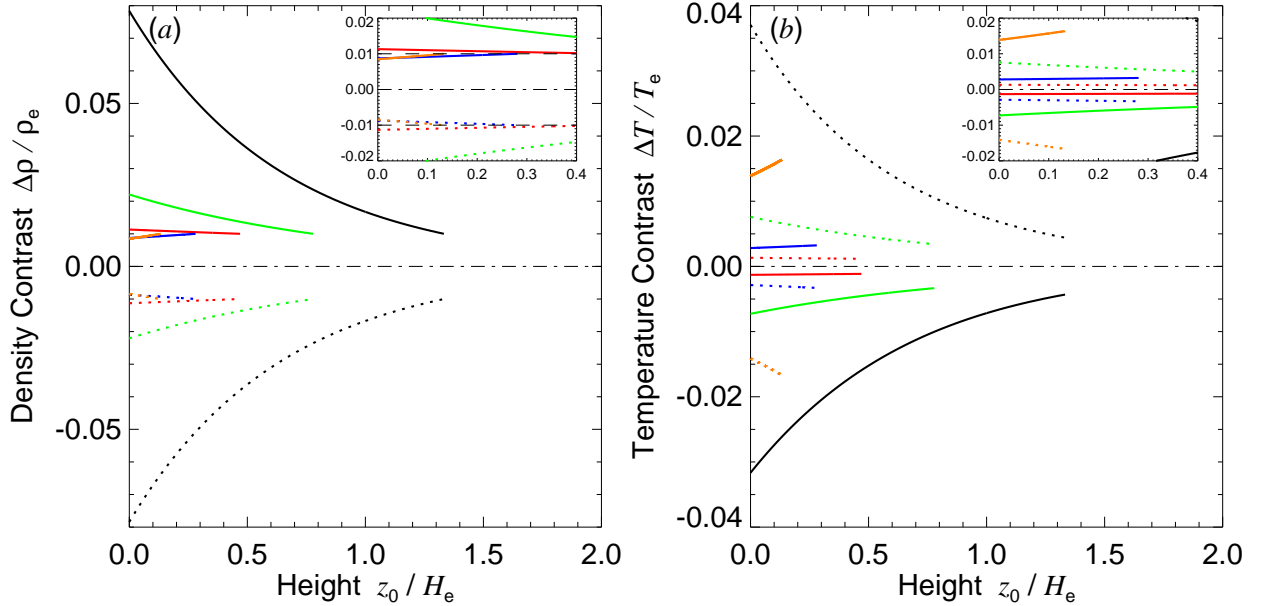}%
        \caption{\small Fractional ($a$) mass density and ($b$) temperature contrast
as a function of height for loops with constant magnetic Bond number and a footpoint
separation of $2X = 250$ Mm. The external atmosphere is isothermal with a scale height
of $H_\e = 75$ Mm. The different colors indicate different Bond numbers with the same
color coding as in Figure 2. For all loops, the fractional density contrast has been
set to a value of $\Delta\rho/\rho_\e = \pm 0.01$ at the loop's apex. The solid curves
are for overdense loops $\Delta\rho >0 $ and the dotted curves are for underdense loops
$\Delta\rho < 0$. Overdense loops correspond to loops that have been heated and undedense
loops to those that have been cooled. Clearly, overdense loops are more physically relevant.
The dot-dashed curve is a reference line for vanishing contrast. The insets display
zoom-in views designed to highlight short loops with small magnetic Bond number. In the
mass density inset, the dashed horizontal lines indicate the values of the contrast fixed
at the apex. The contrasts decrease in magnitude with height for loops with $H_\e < h$
and increase in magnitude for loops with $H_\e > h$, where $h$ is the scale height for
the contrast variables. For the illustrated values of the magnetic Bond number,
$h$ ranges from 34 Mm to 160 Mm,
with the black curve ($\varepsilon = -0.47 \, \pi$) having the smallest value and the orange
curve ($\varepsilon = -0.1 \, \pi$) the largest. The blue curve has $h = 1.06~H_\e$,
thus sitting near the transition between the two behaviors.
\label{fig:TandD_Bond}}%

\end{figure*}%


\begin{figure*}%
        \epsscale{1.0}%
        \plotone{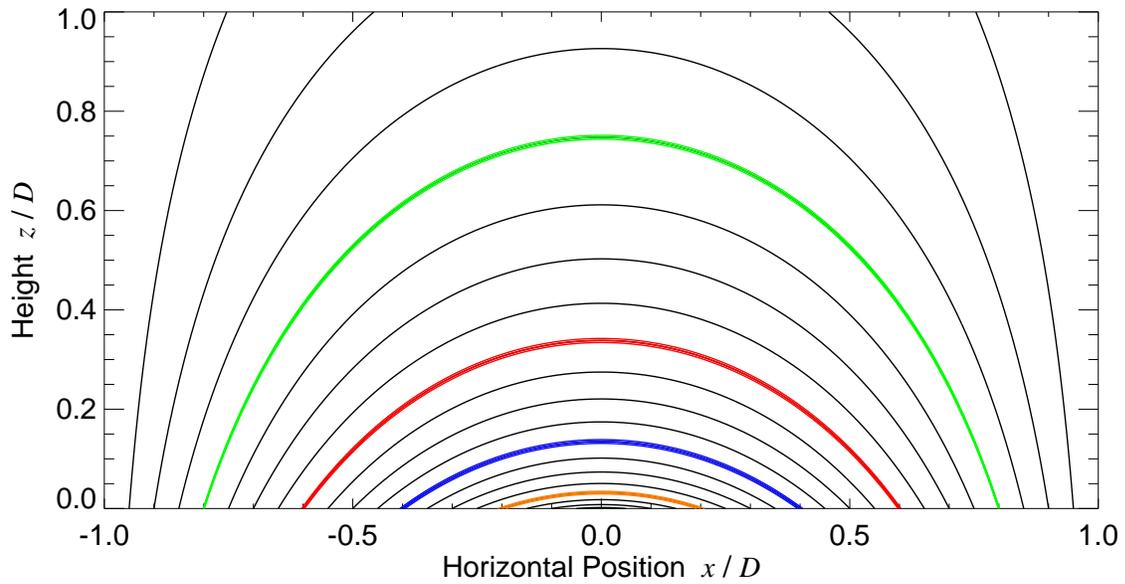}%
        \caption{\small Field lines for a potential corona with embedded magnetic
fibrils, all with constant magnetic Bond number. The coronal field (black lines)
was constructed from a harmonic solution, with a single wavenumber, $\kappa = \pi/(2D)$,
in the $x$-direction---see equation~\eqnref{eqn:Psi}. The colored field lines indicate
magnetic fibrils with different magnetic Bond numbers with the same color coding as
Figure 2 (green: -0.4 $\pi$, red: -0.3 $\pi$, blue: -0.2 $\pi$, and orange: -0.1 $\pi$).
\label{fig:ConstBondField}}%

\end{figure*}%


\begin{figure*}%
        \epsscale{1.0}%
        \plotone{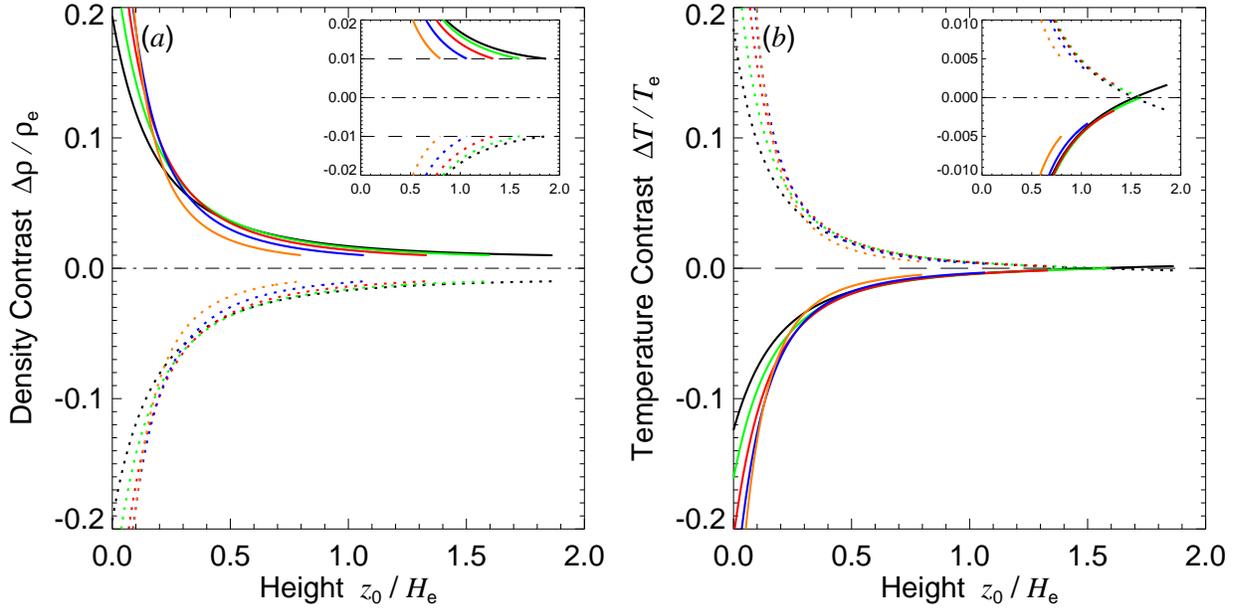}%
        \caption{\small Fractional ($a$) mass density and ($b$) temperature contrast
as a function of height for loops with constant radius of curvature. The external
atmosphere is isothermal with a scale height of $H_\e = 75$ Mm. The different colors
indicate loops with different radii of curvature (black: $R = 175$ Mm, green: 150 Mm,
red: 125 Mm, blue: 100 Mm, orange: 75 Mm). For all loops, the fractional density contrast
has been set to a value of $\Delta\rho/\rho_\e = \pm 0.01$ at the loop's apex. The solid
curves are for overdense loops $\Delta\rho >0 $ (heated loops) and the dotted curves are for underdense
loops $\Delta\rho < 0$ (cooled loops). The dot-dashed curve is a reference line for vanishing contrast
and the dashed curves in the inset correspond to the apex values. The contrasts generally
decrease in magnitude with height. However, loops of sufficient size, $R > 2H_\e$, can have
a temperature inversion at their apex (see the inset in panel $b$).
\label{fig:TandD_circ}}%

\end{figure*}%


\begin{figure*}%
        \epsscale{1.0}%
        \plotone{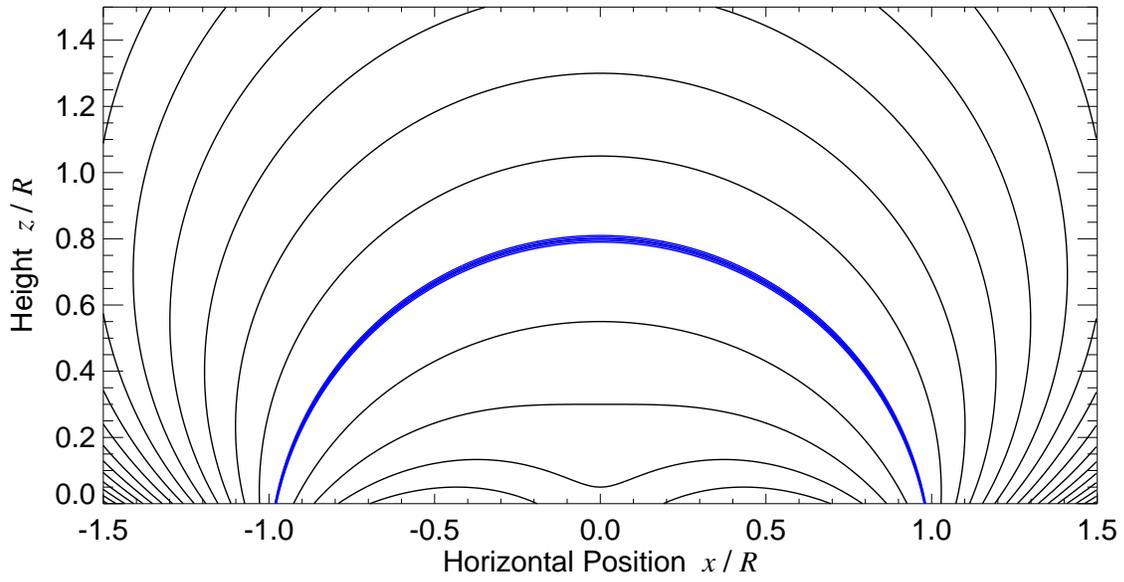}%
        \caption{\small Field lines for a potential corona with an embedded magnetic
fibril with a constant radius of curvature $R$. The field lines spread with height.
\label{fig:ConstCurvField}}%

\end{figure*}%


\begin{thebibliography}{}

\bibitem[Aschwanden(2005)]{Aschwanden:2005}
Aschwanden M.~J. 2005, Physics of the Solar Corona, (Springer-Verlag: Heidelberg), pp. 321--322

\bibitem[Aschwanden(2011)]{Aschwanden:2011-LRSP}
Aschwanden M.~J. 2011, LRSP, 8, 5

\bibitem[Aschwanden(2013)]{Aschwanden:2013}
Aschwanden M.~J. 2013, \apj, 763, 115

\bibitem[Aschwanden et al.(1999)]{Aschwanden:1999}
Aschwanden M.~J., Fletcher, L., Schrijver, C.~J., \& Alexander, D. 1999, \apj, 520, 880

\bibitem[Aschwanden \& Schrijver(1999)]{Aschwanden:2011-apj}
Aschwanden M.~J. \& Schrijver, C.~J. 2011, \apj, 736, 102

\bibitem[Aschwanden et al.(2001)]{Aschwanden:2001}
Aschwanden M.~J., Schrijver, C.~J., Alexander, D. 2001, \apj, 550, 1036

\bibitem[Brooks et al.(2013)]{Brooks:2013}
Brooks, D. H., Warren, H. P., Ugarte-Urra, I., \& Winebarger, A. R. 2013, \apjl, 772, 19

\bibitem[Cargill \& Klimchuk(2004)]{Cargill:2004}
Cargill, P.~J. \& Klimchuk, J.~A. 2004, \apj, 605, 911

\bibitem[Chandrasekhar(1961)]{Chandrasekhar:1961}
Chandrasekhar, S. 1961, Hydrodynamic and Hydromagnetic Stability, (Dover: New York), pp. 464--466

\bibitem[Cheng(1992)]{Cheng:1992}
Cheng, J. 1992, \aap, 264, 243

\bibitem[Choudhuri(1990)]{Choudhuri:1990}
Choudhuri, A.~R. 1990, \aap, 239, 335

\bibitem[DeRosa et al.(2009)]{DeRosa:2009}
DeRosa, M.~L., Schrijver, C.~J., Barnes, G., et al. 2009, \apj, 696, 1780

\bibitem[Jain \& Hindman(2012)]{Jain:2012}
Jain, R. \& Hindman, B.~W. 2012, \aap, 545, 138

\bibitem[Metcalf et al.(1995)]{Metcalf:1995}
Metcalf, T.~R., Jiao, L., Uitenbroek, H., McClymont, A.~N., \& Canfield, R.~C. 1995, \apj, 439, 474

\bibitem[Nakariakov et al.(1999)]{Nakariakov:1999}
Nakariakov, V., Ofman, L., DeLuca, E., Roberts, B., Davila, J.~M. 1999,  Science, 285, 862

\bibitem[Patsourakos \& Klimchuk(2008)]{Patsourakos:2008}
Patsourakos, S. \& Klimchuk, J.~A. 2008, \apj, 689, 1406

\bibitem[Peter et al.(2013)]{Peter:2013}
Peter, H., Bingert, S., Klimchuk, J. A., de Forest, C., Cirtain, J. W., Golub, L.,
Winebarger, A. R., Kobayashi, K., \& Korreck, E. 2013, \aap, submitted, arXiv:1306:4685v1

\bibitem[Reale(2010)]{Reale:2010}
Reale, F. 2010, LRSP, 7,5

\bibitem[Reale(2002)]{Reale:2002}
Reale, F. 2002, \apj, 580, 566

\bibitem[Spruit(1981)]{Spruit:1981}
Spruit, H.~C. 1981, \aap, 98, 155

\bibitem[Verwichte et al.(2004)]{Verwichte:2004}
Verwichte, E., Nakariakov, V.~M., Ofman, L., \& Deluca, E.~E. 2004, \solphys, 223, 77

\bibitem[Wiegelmann et al.(2008)]{Wiegelmann:2009}
Wiegelmann, T., Inhester, B., \& Feng, L. 2009, AnGeo, 27, 2925

\bibitem[Winebarger \& Warren(2004)]{Winebarger:2004}
Winebarger, A.~R. \& Warren, H.~P. 2004, \apj, 610, L129


\end{thebibliography}
\end{document}